\documentclass[aps,pra,reprint,showpacs,groupedaddress]{revtex4-1}
\usepackage[dvips]{graphicx}
\usepackage[dvips,colorlinks=true,bookmarks=false,citecolor=blue,urlcolor=blue]{hyperref} 
\usepackage{mathtools}

\begin{document}
\title{Universal Optimal Estimation of the Polarization of Light with Arbitrary Photon Statistics}

\author{Lu \surname{Zhang}}
\email[Email: ]{lu@ou.edu}
\author{Kam Wai Clifford \surname{Chan}}
\author{Pramode K. \surname{Verma}}
\affiliation{School of Electrical and Computer Engineering, University of Oklahoma--Tulsa, Tulsa, Oklahoma 74135, USA}

\begin{abstract}
A universal and optimal method for the polarimetry of light with arbitrary photon statistics is presented. The method is based on the continuous maximum-likelihood positive operator-valued measure (ML-POVM) for pure polarization states over the surface of the Bloch sphere.  The success probability and the mean fidelity are used as the figures of merit to show its performance. The POVM is found to attain the collective bound of polarization estimation with respect to the mean fidelity.  As demonstrations, explicit results for the $N$ photon Fock state, the phase-randomized coherent state (Poisson distribution), and the thermal light are obtained.  It is found that the estimation performances for the Fock state and the Poisson distribution are almost identical, while that for the thermal light is much worse.  This suggests that thermal light leaks less information to an eavesdropper and hence could potentially provide more security in polarization-encoded quantum communication protocols than a single-mode laser beam as customarily considered.  Finally, comparisons against an optimal adaptive measurement with classical communications are made to show the better and more stable performance of the continuous ML-POVM.
\end{abstract}
\date{\today}

\pacs{03.67.Hk, 03.65.Wj, 03.65.Ta}
\maketitle

\section{Introduction}
The polarization of light is an important resource that has widespread applications.  It provides additional information of the subjects to be inspected in remote sensing~\cite{Egan1992} and microscopy~\cite{McCrone-etal1978}.  It is also used for encoding information in the most advanced quantum key distribution (QKD) protocols~\cite{FerreiraDaSilva-etal2013,Tang-etal2014}.  In fact, the first QKD protocol--BB84--was proposed using the photon polarization as the information carrier~\cite{Bennett-Brassard1984}.

In classical optics, a common method of determining the polarization state of a light beam is by measuring its Stokes parameters~\cite{Berry-etal1977,Chipman1995,Foreman-etal2015}, which is performed by measuring the intensity of the beam in a few fixed measurement bases of the polarization.  To achieve high accuracy, the light beam should have a sufficient number of photons.  For those applications such as quantum communication and quantum imaging~\cite{Israel-etal2014} that operate in the low photon regime, one needs to seek for more efficient approaches as the Stokes parameters measurement is known to be non-optimal~\cite{Bagan-etal2002}.

The problem of the optimal estimation of a two-dimensional quantum state (a qubit) like the photon polarization has been studied extensively in the past two decades~\cite{Jones1994,Massar-Popescu1995,Derka-etal1998,Gill-Massar2000, Bagan-etal2002,Embacher-Narnhofer2004,deBurgh-etal2008,Guta-etal2008,
Fischer-etal2000,Hannemann-etal2002,Bagan-etal2005,Happ-Freybergerl2008,Kravtsov-etal2013}. In those studies, the mean fidelity $F$~\cite{Nielsen-Chuang2000} is commonly chosen as a figure of merit for optimality. It has been shown that, given $N$ copies of a photon with an unknown polarization $\mathbf{r}_0$ picked from the surface of the Bloch sphere, the optimal mean fidelity between the estimate and $\mathbf{r}_0$ is $F_\text{opt} = (N+1)/(N+2)$.  This result can be achieved by collective measurements~\cite{Massar-Popescu1995,Derka-etal1998} or asymptotically by local measurements with classical communications~\cite{Fischer-etal2000,Hannemann-etal2002,Bagan-etal2005,Happ-Freybergerl2008,Kravtsov-etal2013}.

On the other hand, for practical optical sensing, imaging, and communication applications, the number of photons of the light beam is usually not known in advance.  Most light sources in fact exhibit a distribution in the photon number.  The decoy-state BB84 protocol even makes use of varying the mean photon number of the light pulses to enhance the security of the coherent state-based BB84 protocol~\cite{Hwang2003,Lo-etal2005}.

In this paper, we present a universal continuous positive operator-valued measure (POVM) for the polarimetry of multi-photon light beam with arbitrary photon statistics.  We find that the POVM is not only optimal by achieving the collective bound of polarization estimation with respect to the mean fidelity, it is also optimal by being the maximum-likelihood (ML) measurement.  It is a multi-photon generalization of the single-photon ML-POVM in~\cite{Shapiro2008}.

We apply the ML-POVM to study the optimal measurement when the light beam is a Fock state, a phase-randomized single-mode laser pulse, and an incoherent light pulse (a thermal beam).  As a demonstration of the performance, we consider the problem of polarization determination where the number of the different polarizations is finite, which has applications in certain multi-photon quantum communication protocols~\cite{Barbosa-etal2003,Yuen2009,Kak2006,Chan-etal2015}.  We calculate the success probability of estimating the true polarization and the mean fidelity, and make a comparison between the presented POVM and an optimal adaptive local measurement with classical communications.

\section{Maximum-likelihood POVM for incoherent mixture of Fock states}\label{sec:sec_ML-POVM}

In the following, we consider the scenario where the photons have a fixed but unknown polarization vector $\mathbf{r}_0$.  Unlike the previous studies~\cite{Jones1994,Massar-Popescu1995,Derka-etal1998,Gill-Massar2000, Bagan-etal2002,Embacher-Narnhofer2004,deBurgh-etal2008,Guta-etal2008,
Fischer-etal2000,Hannemann-etal2002,Bagan-etal2005,Happ-Freybergerl2008,Kravtsov-etal2013} that represent the quantum state as a tensor product of $N$ copies of a polarized single photon, we utilize the Fock basis to represent the multi-photon state.  These two approaches are equivalent to each other, but the latter method provides a more concise notation when the number of photons is indefinite.

In the Fock basis, a polarized single photon is given by $|1\rangle_\mathbf{r} = a_\mathbf{r}^\dagger |0\rangle$, where
\begin{equation}
    a_\mathbf{r}^\dagger
    =
    \cos \frac{\theta}{2} a_H^\dagger + e^{i\phi} \sin \frac{\theta}{2} a_V^\dagger,
    \label{mode_operator}
\end{equation}
in which $\theta$ and $\phi$ are the spherical coordinates of the polarization vector $\mathbf{r}$ on the Bloch sphere, and $a_H$ and $a_V$ are the annihilation operators for the north pole and the south pole, which we designate as the horizontal and vertical polarizations respectively.  The operator $a_\mathbf{r}$ satisfies the commutation relation
\begin{equation}
    [a_\mathbf{r}, a_{\mathbf{r}'}^\dagger] = {}_\mathbf{r}\langle 1|1 \rangle_{\mathbf{r}'} \equiv f_{\mathbf{r}\mathbf{r}'}.
    \label{eq:commutation_relation}
\end{equation}
Note that $\left|f_{\mathbf{r}\mathbf{r}'}\right|^2$ is the fidelity between two pure qubits with polarizations $\mathbf{r}$ and $\mathbf{r}'$.
The $n$ photon Fock state basis is then produced by applying the creation operator $a_\mathbf{r}^{\dagger}$ successively, i.e.,
\begin{equation}
    |n\rangle_\mathbf{r} = \frac{a_\mathbf{r}^{\dagger n}}{\sqrt{n!}} |0\rangle.
    \label{eq:n_photon_state}
\end{equation}
It can be shown that
${}_\mathbf{r}\langle n | m \rangle_{\mathbf{r}'} =  f_{\mathbf{r}\mathbf{r}'}^n \delta_{nm}$,
where $\delta_{nm}$ is the Kronecker delta.

In the following analysis, we focus on a light beam with certain photon statistics specified by the photon number distribution $P_n$ instead of a coherent superposition of the $n$ photon state~(\ref{eq:n_photon_state}).
The quantum state in this case is written as
\begin{equation}
    \rho(\mathbf{r})
    =
    \sum_{n=0}^{\infty} P_n |n \rangle_\mathbf{r}\langle n|.
    \label{eq:incoherent mixture of Fock states}
\end{equation}
This is motivated by the fact that the phase information is usually dismissed intentionally in quantum communication to increase security~\cite{Zhao-etal2007,Tang-etal2013}, or the phase is simply unavailable as in incoherent imaging and remote sensing that operate in the photon-counting mode.  In addition, this simplification allows us to write the optimal POVM explicitly.  It should be noted that $\rho(\mathbf{r})$ is a mixed state in the photon number, not in the polarization as considered in~\cite{Bagan-etal2004}.

To quantify the optimality of the measurement, the fidelity~\cite{Nielsen-Chuang2000} is usually chosen as the figure of merit for the purpose of quantum state tomography.  On the other hand, in quantum communication, one is more concerned about the error of estimating the correct bit value.  In this respect, the maximum-likelihood measurement provides a more natural choice.  These two quantities nevertheless are connected as to be discussed in the next section.

According to quantum estimation theory~\cite{Helstrom1976}, the maximum-likelihood POVM $\Pi(\mathbf{r})$ satisfies the following conditions:
\begin{equation}
    \left[\Upsilon - W(\mathbf{r}) \right] \Pi(\mathbf{r}) = \Pi(\mathbf{r}) \left[\Upsilon - W(\mathbf{r}) \right] = 0,
    \label{condition1}
\end{equation}
and
\begin{equation}
    \Upsilon - W(\mathbf{r}) \geq 0,
    \label{condition2}
\end{equation}
where
\begin{equation}
    W(\mathbf{r})
    \equiv
    \int_S d\mathbf{r}_0 \ p(\mathbf{r}_0) C(\mathbf{r}, \mathbf{r}_0) \rho(\mathbf{r}_0)
    = \frac{\rho(\mathbf{r})}{4\pi}
    \label{W_function}
\end{equation}
is the Hermitian risk operator with a uniform prior distribution $p(\mathbf{r}_0) = 1/4\pi$ and a delta cost function $C(\mathbf{r}, \mathbf{r}_0) = \delta(\mathbf{r}-\mathbf{r}_0)$.  Here $\Upsilon$ is an Hermitian Lagrange operator defined by
\begin{equation}
    \Upsilon
    \equiv
    \int_S d\mathbf{r} \ W(\mathbf{r}) \Pi(\mathbf{r}) .
    \label{gamma_function}
\end{equation}
Note that the integration is over the Bloch surface $S$ with $d\mathbf{r} = \sin\theta d\theta d\phi$.
We find that the operator
\begin{equation}
    \Pi(\mathbf{r})
    =
    \sum_{n=0}^{\infty} \frac{n+1}{4\pi} |{n}\rangle_\mathbf{r} \langle{n}|
    \label{POVM}
\end{equation}
satisfies the conditions~(\ref{condition1}) and~(\ref{condition2}).

To see that, first of all, it should be noted that Eq.~(\ref{POVM}) forms a legitimate continuous POVM (in $\mathbf{r}$), \textit{viz}., $\Pi(\mathbf{r}) > 0$ and
\begin{equation}
    \int_S \Pi(\mathbf{r}) d\mathbf{r}
    =
    I,
    \label{completeness}
\end{equation}
where
\begin{eqnarray}
    I
    &\equiv&
    \sum_{n=0}^{\infty} I_n
    \equiv
    \sum_{n=0}^{\infty}
    \left[\sum_{m=0}^{n}|{m}\rangle_H \langle{m}| \otimes |{n-m}\rangle_V \langle{n-m}|\right]
\nonumber\\
    &=&
    \sum_{n=0}^{\infty}|n\rangle_H \langle n| \otimes \sum_{m=0}^{\infty}|m \rangle_V \langle m|
    \label{identity}
\end{eqnarray}
is the identity operator of the infinite dimensional Fock space and $I_n$ is the identity operator of the $n$ photon subspace.  In addition,
substituting Eqs.~(\ref{W_function}) and (\ref{POVM}) into Eq.~(\ref{gamma_function}), we obtain
\begin{equation}
    \Upsilon=\frac{1}{4\pi} \sum_{n=0}^{\infty} P_n I_n ,
    \label{gamma_expression}
\end{equation}
which is Hermitian. Then Eq.~(\ref{condition1}) is easily verified by substituting Eqs.~(\ref{W_function}),~(\ref{POVM}) and (\ref{gamma_expression}) into Eq.~(\ref{condition1}). To prove Eq.~(\ref{condition2}), we first organize the operator $\Upsilon - W(\mathbf{r})$ into a more suggestive form:
\begin{equation}
    \Upsilon - W(\mathbf{r})
    =
    \frac{1}{4\pi} \sum_{n=0}^{\infty} P_n \left(I_n-|{n}\rangle_\mathbf{r}\langle{n}|\otimes |{0}\rangle_{-\mathbf{r}}\langle{0}|\right) ,
    \label{temp1}
\end{equation}
where the polarization $-\mathbf{r}$ is perpendicular to $\mathbf{r}$. Now $I_n$ can be expanded in any orthogonal polarization bases:
\begin{equation}
    I_n
    =
    \sum_{m=0}^{n} |{m}\rangle_\mathbf{r}\langle{m}| \otimes |{n-m}\rangle_{-\mathbf{r}}\langle{n-m}| .
    \label{temp2}
\end{equation}
Therefore $\Upsilon - W(\mathbf{r})$ is a non-negative definite operator.

\section{Analysis of the continuous maximum-likelihood POVM}

\begin{figure}[t!]
    \centering
    \includegraphics[width=8.6cm]{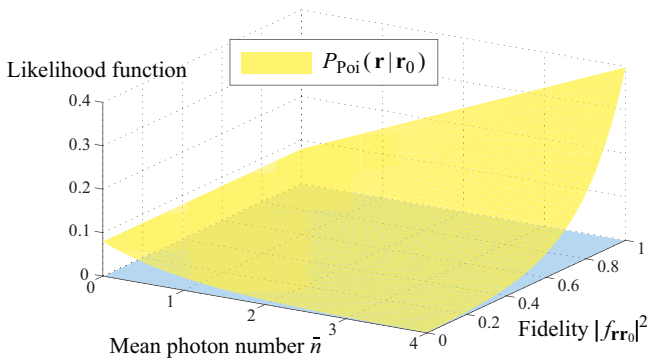}
    \caption{(Color online) Plot of the likelihood $P_\text{Poi}(\mathbf{r} | \mathbf{r}_0)$ of the Poisson distribution as a function of the mean photon number $\bar{n}$ and the fidelity $|f_{\mathbf{r}\mathbf{r}_0}|^2$ (upper yellow surface). The bottom blue surface denotes the plane of zero probability.}
    \label{fig:pdf_mu_f}
\end{figure}

The maximum-likelihood estimate of the initial polarization $\mathbf{r}_0$ given the POVM polarimeter's output $\mathbf{r}$ is
\begin{equation}
    \mathbf{r}_{0_\text{ML}}
    = \arg\max_{\mathbf{r}\in S} P(\mathbf{r} | \mathbf{r}_0)
    = \mathbf{r},
\end{equation}
where $P(\mathbf{r} | \mathbf{r}_0)$ is the likelihood function with the ML-POVM:
\begin{equation}
    P(\mathbf{r} | \mathbf{r}_0)
    =
    \text{Tr}\left[\Pi(\mathbf{r}) \rho(\mathbf{r}_0)\right]
    =
    \sum_{n=0}^\infty \frac{n+1}{4\pi} P_n \left|f_{\mathbf{r}\mathbf{r}_0}\right|^{2n} .
    \label{eq:conditional_prob}
\end{equation}
As mentioned earlier, $\left|f_{\mathbf{r}\mathbf{r}_0}\right|^2 = \frac{1}{2}\left(1+\mathbf{r}\cdot\mathbf{r}_0\right)$ is the fidelity between the polarizations $\mathbf{r}$ and $\mathbf{r}_0$.

Certain cases of the given prior photon distribution $P_n$ admit closed-form expressions for the likelihood function.  For example, the $N$ photon Fock state with $P_{n,N} = \delta_{nN}$ gives $P_N(\mathbf{r} | \mathbf{r}_0) = (N+1)\left|f_{\mathbf{r}\mathbf{r}_0}\right|^{2N}/4\pi$, the Poisson distribution $P_{n,\text{Poi}} = e^{-\bar{n}}\bar{n}^n/n!$ with a mean photon number $\bar{n}$ describing a phase-randomized single-mode laser beam gives
\begin{equation}
    P_\text{Poi}(\mathbf{r} | \mathbf{r}_0)
    =
    \frac{e^{-\bar{n}\left(1-\left|f_{\mathbf{r}\mathbf{r}_0}\right|^2\right)}}{4\pi}
    \left(1 + \bar{n}\left|f_{\mathbf{r}\mathbf{r}_0}\right|^2 \right) ,
    \label{pdf}
\end{equation}
and the Bose-Einstein distribution $P_{n,\text{th}} = \bar{n}^n/(1+\bar{n})^{n+1}$ modeling a thermal beam gives
\begin{equation}
    P_\text{th}(\mathbf{r} | \mathbf{r}_0)
    =
    \frac{1}{4\pi(1+\bar{n})\left(1-\frac{\bar{n}}{1+\bar{n}}\left|f_{\mathbf{r}\mathbf{r}_0}\right|^2\right)^2} .
\end{equation}

As expected, the maxima of these likelihood functions occur at $|f_{\mathbf{r}\mathbf{r}_0}|^2 = 1$ when $\mathbf{r} = \mathbf{r}_0$.  Moreover, the likelihoods behave like the delta function $\delta(1-\left|f_{\mathbf{r}\mathbf{r}_0}\right|^2)$ in the large (mean) photon number limit.  On the other hand, when the (mean) photon number tends to zero, the probability approaches the constant $1/4\pi$ independent of the fidelity, which corresponds to the scenario the same as a random guess.
As an illustration, a plot of the likelihood function for the Poisson distribution against the mean photon number $\bar{n}$ and fidelity $\left|f_{\mathbf{r}\mathbf{r}_0}\right|^{2}$ is shown in Fig.~\ref{fig:pdf_mu_f}.

For quantum communication applications such as the $\alpha\eta$ protocol~\cite{Barbosa-etal2003}, Y00 protocol~\cite{Yuen2009} and the three-stage protocol~\cite{Kak2006,Chan-etal2015} that utilize the polarization states in different bases to encrypt the bit values, it is useful to determine the optimal success probability of estimating the polarization when the knowledge of the basis is absent in order to bound the potential information leak to the eavesdropper.  Since the polarization states on the Bloch sphere is continuous, we need to define a finite region for any polarization vector so as to obtain a nonzero estimation probability. Here we choose the finite region $S_\epsilon(\mathbf{r}_0)$ in the form of a circle on $S$ around $\mathbf{r}_0$ with an angular diameter $2\epsilon$ where $\epsilon \in [0,\pi]$. Therefore, the success probability reads as
\begin{equation}
    Q(\epsilon)
    \equiv
    \int_{S_\epsilon(\mathbf{r}_0)} \hspace{-5mm} d\mathbf{r} \ P(\mathbf{r}|\mathbf{r}_0)
    =
    1 - \sum_{n=0}^\infty P_n \left(\frac{1+\cos\epsilon}{2}\right)^{n+1},
    \label{eq:P_s}
\end{equation}
where the factor $(1+\cos\epsilon)/2$ is the fidelity of a qubit with a neighboring qubit at an angle $\epsilon$ away. We then obtain the closed forms for the Fock state, the Poisson distribution, and the thermal distribution:
\begin{equation}
    Q_N(\epsilon)
    =
    1 - \left(\frac{1+\cos\epsilon}{2}\right)^{N+1}
    \approx
    1 - e^{-\frac{\epsilon^2}{4}(N+1)} ,
\end{equation}
\begin{equation}
    Q_{\text{Poi}}(\epsilon)
    =
    1 - \frac{1+\cos\epsilon}{2} e^{-\frac{\bar{n}}{2}\left(1-\cos\epsilon\right)}
    \approx
    1 - e^{-\frac{\epsilon^2}{4}(\bar{n}+1)} ,
\end{equation}
and
\begin{equation}
    Q_{\text{th}}(\epsilon)
    =
    1 - \frac{1+\cos\epsilon}{2+\bar{n}(1-\cos\epsilon)}
    \approx
    1 - \frac{4-\epsilon^2}{4+\epsilon^2 \bar{n}} ,
\end{equation}
where the approximations correspond to the small $\epsilon$ limit.

\begin{figure}[t!]
    \centering
    \includegraphics[height=6.6cm]{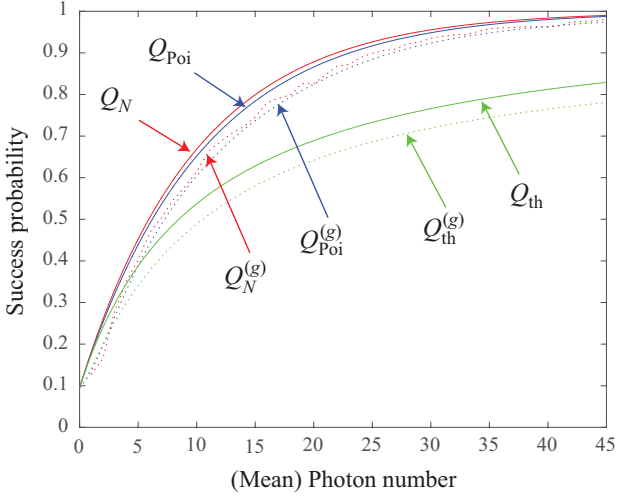}
    \caption{(Color online) Plots of the success probabilities $Q(\epsilon)$ when $\epsilon =0.2\pi$ for the Fock state ($Q_N$, red), the Poisson distribution ($Q_{\text{Poi}}$, blue), and the thermal distribution ($Q_{\text{th}}$, green), using the ML-POVM (solid curves).  The quantities $Q_N^{(g)}$, $Q_{\text{Poi}}^{(g)}$ and $Q_{\text{th}}^{(g)}$ are the corresponding success probabilities calculated using the greedy scheme (dotted curves). Note that the curves for the Fock state and the Poisson distribution are almost indistinguishable.  For $Q_N$ and $Q_N^{(g)}$, the discrete points are interpolated to give better visualization.}
    \label{fig:success_probability_analytic}
\end{figure}

The success probabilities $Q(\epsilon)$ when $\epsilon =0.2\pi$ are shown in Fig.~\ref{fig:success_probability_analytic} (solid curves).
It should be noted that the error probability $1-Q_N$ for the Fock state exhibits the exponential decay of the quantum Chernoff bound for the discrimination of two quantum states~\cite{Audenaert-etal2007}.
Interestingly, the success probability of the Poisson distribution bears the same scaling as that of the Fock state.  For the thermal distribution, its error probability decreases as $\bar{n}^{-1}$, which is much slower than the other two cases.  From the perspective of quantum communication protocols in~\cite{Barbosa-etal2003,Yuen2009,Kak2006,Chan-etal2015}, these results suggest that polarized light from a thermal source potentially is more secure than a single-mode laser beam as customarily considered. It is noticeable that the success probability has a non-zero value $(1-\cos{\epsilon})/2 \approx 0.095$ for the vacuum state scenario. This is because the finite region defined for a success estimate gives a non-zero success probability even for a random guess.

To compare with the results of previous studies, we also calculate the mean fidelity using the continuous ML-POVM, which is given by
\begin{equation}
    F
    =
    \int_S d\mathbf{r} \, d\mathbf{r}_0 \left|f_{\mathbf{r}_0\mathbf{r}_{0_\text{ML}}}\right|^2 P(\mathbf{r} | \mathbf{r}_0) p(\mathbf{r}_0)
    =
    \sum_{n=0}^\infty \frac{n+1}{n+2} P_n .
    \label{eq:Fidelity}
\end{equation}
This is the same as the bound obtained by collective measurements for arbitrary photon statistics. The mean fidelity for the Fock state is then $F_N=(N+1)/(N+2)$. For the Poisson and thermal distributions, we get
\begin{equation}
    F_\text{Poi}(\bar{n})
    =
    \frac{1 - \bar{n} + \bar{n}^2 - e^{-\bar{n}}}{\bar{n}^2}
    \approx
    1 - \frac{1}{\bar{n}} ,
\end{equation}
and
\begin{equation}
    F_\text{th}(\bar{n}) = \frac{(1 + \bar{n}) (\bar{n} - \log(1+\bar{n}))}{\bar{n}^2}
    \approx
    1 - \frac{\log(1+\bar{n})}{\bar{n}} ,
\end{equation}
where the approximations correspond to the large mean photon number $\bar{n}$ limit.
\begin{figure}[!t]
    \centering
    \includegraphics[height=6.6cm]{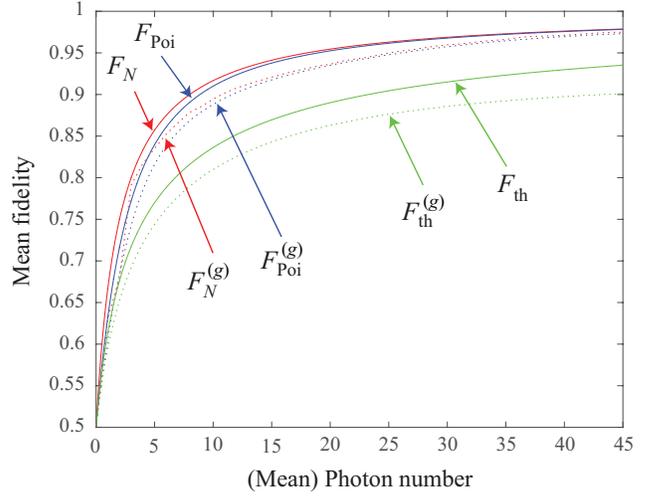}
    \caption{(Color online) Plots of the mean fidelities for the Fock state ($F_N$, red), the Poisson distribution ($F_\text{Poi}$, blue), and the thermal distribution ($F_\text{th}$, green), using the ML-POVM (solid curves).  The quantities $F_N^{(g)}$, $F_{\text{Poi}}^{(g)}$, and $F_{\text{th}}^{(g)}$ are the corresponding mean fidelities calculated using the greedy scheme (dotted curves).  For $F_N$ and $F_N^{(g)}$, the discrete points are interpolated to give better visualization.}
    \label{fig:fidelity_comparison}
\end{figure}
Figure~\ref{fig:fidelity_comparison} shows the plots of the mean fidelities under the three scenarios (solid curves).  Similar to the success probability, the mean fidelity of the Poisson distribution approaches that of the Fock state very quickly, while it increases at a much slower rate for the thermal distribution.

As mentioned previously, the success probability and the mean fidelity are connected.  This can be seen from the definition of the successful region $S_\epsilon(\mathbf{r}_0)$ which includes all the polarization vectors $\mathbf{r}$ with a fidelity $\left|f_{\mathbf{r}\mathbf{r}_0}\right|^2 \le \frac{1}{2}\left(1 + \cos\epsilon\right)$.  Therefore the mean fidelity gives the mean performance while the success probability gives the point-wise performance.  In addition, Eq.~(\ref{eq:conditional_prob}) is essentially a function of the fidelity.  In fact, the likelihood can be shown to be proportional to the probability density function of the fidelity.

\section{Comparison Between the ML-POVM and an Optimal Adaptive Measurement with Classical Communications}

In this section, we compare the performances of the presented ML-POVM against an optimal measurement, the greedy scheme~\cite{Bagan-etal2005}, with respect to the success probability and the fidelity. The greedy scheme utilizes an adaptive local measurement with classical communications to measure the polarization state given $n$ copies of a polarized single photon.  The classical communications is reflected by the fact that the current operations can depend on the previous steps.

The greedy scheme works in this way: for the first three photons, the measurement bases are taken to be the $\mathbf{x}$, $\mathbf{y}$ and $\mathbf{z}$ directions.  For the $k^\text{th}$ measurement with $k \ge 4$, the measurement basis is chosen to be the direction $\mathbf{m}_k$, which is found by maximizing the mean fidelity of the current step based on the measurement results of the $(k-1)^\text{th}$ step (see \cite{Bagan-etal2005} for detail).  The measurement outcome through the $n$ steps is denoted by $\chi_n = i_n i_{n-1} \cdots i_2 i_1$ where $i_k$ is $k^\text{th}$ outcome with $i_k = 0$ if the photon detector in the direction $\mathbf{m}_k$ clicks and $i_k = 1$ if the detector in the orthogonal direction clicks.  The likelihood function after the $n$-step measurements reads as
\begin{equation}
    P^{(g)}(\chi_n|\mathbf{r}_0)
    =
    \displaystyle \prod_{k=1}^{n}\frac{1+\mathbf{m}_k\cdot \mathbf{r}_0}{2}=\displaystyle \prod_{k=1}^{n}\left|f_{\mathbf{m}_k\mathbf{r}_0}\right|^2 .
    \label{eq:likelihood_greedy}
\end{equation}
Then the optimal estimate of the polarization $\mathbf{r}_0$ given the $n$-step greedy scheme measurement results $\chi_n$ is
\begin{equation}
    \mathbf{r}_0^{(g)}
    = \arg\max_{\mathbf{r}\in S} F_n^{(g)}
    =\frac{\mathbf{V}(\chi_n)}{|\mathbf{V}(\chi_n)|},
    \label{eq:OptimalEstimate_greedy}
\end{equation}
where
\begin{equation}
    \mathbf{V}(\chi_n)
    =
    \int d\mathbf{r}_0\ P^{(g)}(\chi_n|\mathbf{r}_0) p(\mathbf{r}_0)\mathbf{r}_0
    \label{eq:V_chi_greedy}
\end{equation}
gives the maximal $n$-photon mean fidelity
\begin{equation}
    F_n^{(g)}
    =
    \frac{1}{2}\left(1+\sum_{\chi_n}|\mathbf{V}(\chi_n)|\right) ,
    \label{eq:Fidelity_greedy}
\end{equation}
in which the summation runs over all the possible values for $\chi_n$.
Note that the measure of mean fidelity for the greedy scheme as in Eq.~(\ref{eq:Fidelity_greedy}) is derived from Eq. (2.3) in \cite{Bagan-etal2005}, which is identical to that for the ML-POVM method as in Eq.~(\ref{eq:Fidelity}). They are both the same as the mean Uhlmann fidelity for pure states.
This method has been proven to be better than fixed-basis measurement schemes, such as the Stokes parameters measurement, and it can approach the collective bound in a quick manner. It is also the best local operation with classical communications one can perform when the photon number is unknown in advance.

It is remarked that the number of the possible measurement directions $\mathbf{m}_k$ for the $k^\text{th}$ measurement is $2^k-1$.  They in principle can be computed in advance so that the measurements can be performed rapidly.  However, when $n$ is large, i.e., $n > 20$, it is quite impractical to perform the computation, and as a consequence, one is unable to obtain a closed-form expression for the mean fidelity based on Eq.~(\ref{eq:likelihood_greedy}).

To proceed, we apply the greedy scheme to the analysis here by assuming that the $n$ photon Fock state can be separated into $n$ individual single photons.  Experimentally, this could in principle be done by using many beam splitters to separate the input Fock state so that the outputs of the beam splitters are either a vacuum state or a single photon state with high probability.  In addition, according to the quantum states represented in Eq.~(\ref{eq:incoherent mixture of Fock states}), one obtains a beam containing $n$ photons with probability $P_n$. Therefore, the mean fidelity is calculated by $F^{(g)}=\sum_{n=0}^{\infty}P_n F_n^{(g)}$.

We have performed numerical simulations using $10^4$ samples.  The success probability and the mean fidelity are plotted in Fig.~\ref{fig:success_probability_analytic} (dotted curves) and Fig.~\ref{fig:fidelity_comparison} (dotted curves) respectively.  The bumps on the curves are due to the randomness in the simulations.  For the Fock state and the Poisson distribution, the two quantities using the greedy scheme are smaller than those using the ML-POVM in the low photon regime. These two methods approach the same performance when the mean number of photons is sufficiently large.
The greedy scheme always performs worse than the ML-POVM because it uses the information of the photons one by one sequentially and optimizes the mean fidelity merely based on the previous immediate measurement.  This is in contrast to the ML-POVM which is a collective measurement that uses the information of all the photons in a single step.
For the thermal distribution, it is seen that the ML-POVM is significantly better than the greedy scheme even when the mean photon number is large.  This can be explained by the fact that, for the thermal distribution, there are always more contributions from the small photon number states than from the large photon number states, and the former tend to lower the mean fidelity.

As mentioned earlier, the mean fidelity only gives the average performance of the polarization estimation.  In actual estimations, the fidelities obtained for different unknown $\mathbf{r}_0$'s can vary with a large range depending on the estimation method as well as the photon number distribution.  Therefore, to demonstrate the stability of the presented methods, we also calculate the variances of the fidelities under different initial polarizations.  The variance of the continuous ML-POVM reads as
\begin{equation}
    \Delta^2 F
    =
    \sum_{n=0}^\infty \frac{n+1}{n+3} P_n - \left(\sum_{n=0}^\infty \frac{n+1}{n+2} P_n\right)^2 .
    \label{eq:std}
\end{equation}
The explicit forms for the Fock state, the Poisson distribution and the thermal distribution are respectively
\begin{equation}
    \Delta^2 F_N
    =
    \frac{N+1}{(N+3)(N+2)^2} ,
    \label{eq:std_fock}
\end{equation}
\begin{equation}
    \Delta^2 F_\text{Poi}(\bar{n})
    =
    \frac{(\bar{n}^2-2\bar{n}-1)+2e^{-\bar{n}}(1+\bar{n}+\bar{n}^2)-e^{-2\bar{n}}}{\bar{n}^4} ,
    \label{eq:std_Poi}
\end{equation}
and
\begin{equation}
    \Delta^2 F_\text{th}(\bar{n})
    =
    \frac{(1+\bar{n})\left(\bar{n}^2-(1+\bar{n})\left(\log{(1+\bar{n})}\right)^2\right)}{\bar{n}^4} .
    \label{eq:std_th}
\end{equation}

\begin{figure}[!t]
    \centering
    \includegraphics[height=18.5cm]{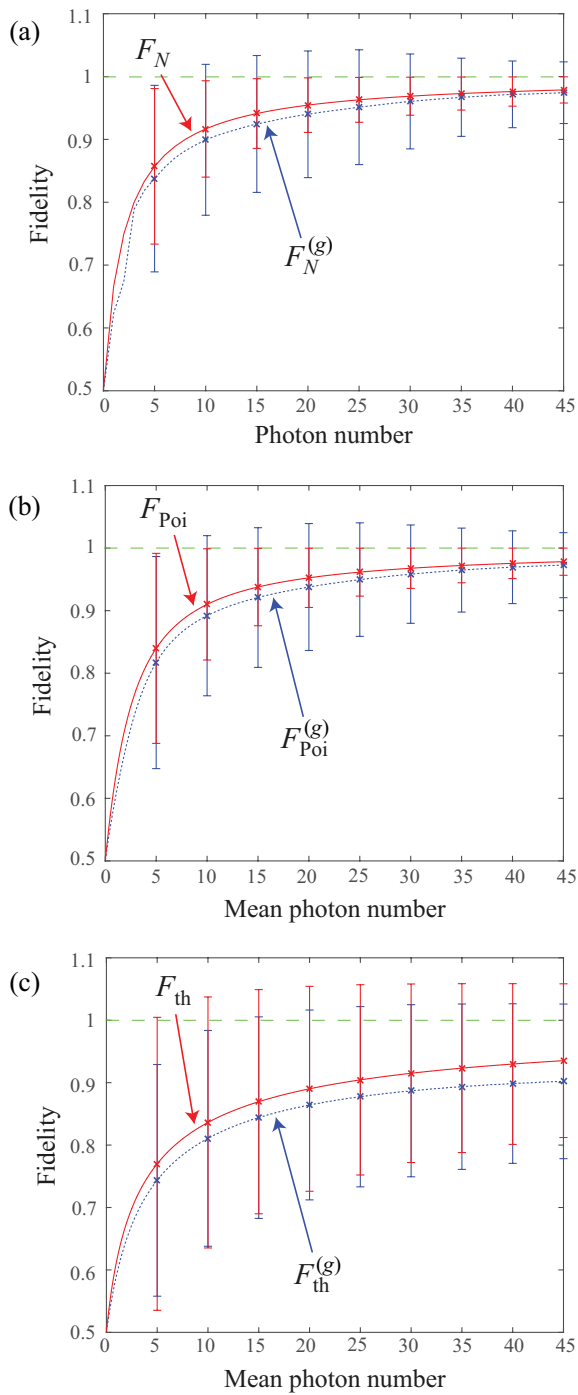}
    \caption{(Color online) Plots of the mean fidelities with standard deviations for (a) the Fock state ($F_N/F_N^{(g)}$), (b) the Poisson distribution ($F_\text{Poi}/F_{\text{Poi}}^{(g)}$), and (c) the thermal distribution ($F_\text{th}/F_{\text{th}}^{(g)}$), using the ML-POVM (red solid curves) and the greedy scheme (blue dotted curves).  Note that only certain points are plotted for the standard deviations.  The green dashed lines are the upper bounds of the fidelity.}
    \label{fig:standard_variance_comparison}
\end{figure}

On the other hand, the variance using the greedy scheme is calculated from the numerical simulations. Figure~\ref{fig:standard_variance_comparison} depicts the plots of the standard deviations (error bars) using the ML-POVM (red solid curves) and the greedy scheme (blue dotted curves).  We have performed the simulations using different number of samples and the standard deviations are found to exhibit only small variations due to the randomness in the sampling.  It is remarked that even though the mean fidelity plus one standard deviation can be greater than one, the actual fidelity found is always between zero and one.  The variance only gives a rough estimate of the range of the possible fidelity values.

From Fig.~\ref{fig:standard_variance_comparison}, it is noticeable that the standard deviation of the ML-POVM scheme is much smaller than that of the greedy scheme for both Fock state and Poisson distribution, and they decrease quickly with the increasing mean photon number.  It suggests that the ML-POVM method is a more stable scheme than the greedy scheme with respect to the fidelity, and the higher the mean photon number is, the more stable the method gets.  On the other hand, the standard deviation for the thermal distribution is much larger than that for the other two scenarios and it decreases at a much slower rate.  These can be shown from Eqs.~(\ref{eq:std_fock})--(\ref{eq:std_th}), where we find $\Delta F_N \sim \Delta F_\text{Poi} \sim \bar{n}^{-1}$ and $\Delta F_\text{th} \sim \bar{n}^{-1/2}$ when the (mean) photon number $\bar{n}$ is large.  Once again this implies that the thermal light can provide higher security for multi-photon polarization-encoded quantum communication protocols.

\section{Summary}

In this paper, a method of measuring any polarization of multi-photon state with arbitrary photon number statistics is investigated. It is achieved by performing a continuous positive operator-valued measure, which is optimal by being a maximum-likelihood measurement.  The likelihood function of the estimate $\mathbf{r}$ is found explicitly given the prior photon distribution. These results provide the computational tools for applications such as the theoretical security analysis of multi-photon quantum communication. In particular, we study the cases of the Fock state, the Poisson distribution, and the thermal distribution in detail in terms of the success probability and the mean fidelity. Surprisingly, the Poisson distribution with mean photon number $N$, which can model a phase-randomized coherent state, performs almost as good as an $N$ photon Fock state. In addition, we find that the thermal distribution always gives a much worse estimate than the other cases.  This suggests that thermal light sources such as light-emitting diodes or a laser beam containing many incoherent modes may achieve a more secure information transmission than the commonly used coherent laser beam in multi-photon quantum communication protocols.

We also compare the ML-POVM against an optimal adaptive local measurement with classical communications (the greedy scheme).  For the cases of Fock state and Poisson distribution, both the success probability and the mean fidelity show larger values for the continuous ML-POVM in the low photon regime.  For the thermal distribution, the ML-POVM is significantly better than the greedy scheme even when the mean photon number is large.  It is also noticeable that the ML-POVM method is a more stable scheme from the perspective of the fluctuation of the estimation fidelity.  Again, the polarization information retrieved from the thermal light is much less stable than the other two cases.

Finally, one may wonder how the continuous ML-POVM could be implemented experimentally.  One possibility would be to follow the discretized polarimeter for the single photon ML-POVM in~\cite{Shapiro2008}, which operates by splitting the single photon into $M$ different paths, and on each of the $M$ output modes, a standard projective polarization analysis is performed.  To extend the scheme to the multi-photon situation as discussed here, the single-photon detector at each of the output mode is simply replaced by a photon-number resolving detector.  However, in order to realize the ML-POVM in Eq.~(\ref{POVM}), only the events of all photons going to one of the paths will be used; the many more other events will have to be discarded, making this scheme not practical.  Nevertheless, since the presented ML-POVM works for arbitrary photon statistics, it may be possible that the discarded events could be retained if some proper post-processing of the data is made by taking into account the beam splitting process.  This will be a topic of further investigation.

\begin{acknowledgments}
The authors thank Guangyu Fang for useful discussions.  This research is supported in part by National Science Foundation Grant No.~1117179.
\end{acknowledgments}



\begin{thebibliography}{9}


\bibitem{Egan1992}
W. G. Egan,
Proc. SPIE 1747, Polarization and Remote Sensing, 2 (December 8, 1992); doi:10.1117/12.142571. 

\bibitem{McCrone-etal1978}
W. C. McCrone, L. B. McCrone, and J. G. Delly,
\textit{Polarized light microscopy}
(Microscope Publications, 1978).

\bibitem{FerreiraDaSilva-etal2013}
T. Ferreira da Silva, D. Vitoreti, G. B. Xavier, G. C. do Amaral, G. P. Tempor\~{a}o, and J. P. von der Weid,
Phys. Rev. A \textbf{88}, 052303 (2013).

\bibitem{Tang-etal2014}
Z. Tang, Z. Liao, F. Xu, B. Qi, L. Qian, and H.-K. Lo,
Phys. Rev. Lett. \textbf{112}, 190503 (2014).

\bibitem{Bennett-Brassard1984}
C. H. Bennett and G. Brassard,
Proceedings of the IEEE Conference on Computers, Systems and Signal Processing (Bangalore, India: IEEE) pp. 175. (1984).

\bibitem{Chipman1995}
R. A. Chipman, in \textit{Handbook of Optics}, 2nd ed., M. Bass, ed. (McGraw-Hill, New York, 1995), Chap. 22.

\bibitem{Berry-etal1977}
H. G. Berry, G. Gabrielse, and A. E. Livingston,
Appl. Opt. \textbf{16}, 3200 (1977).

\bibitem{Foreman-etal2015}
M. R. Foreman, A. Favaro, and A. Aiello,
Phys. Rev. Lett. \textbf{115}, 263901 (2015).

\bibitem{Israel-etal2014}
Y. Israel, S. Rosen, and Y. Silberberg,
Phys. Rev. Lett. \textbf{112}, 103604 (2014).

\bibitem{Bagan-etal2002}
E. Bagan, M. Baig, and R. Mu\~{n}oz-Tapia,
Phys. Rev. Lett. \textbf{89}, 277904 (2002).

\bibitem{Jones1994}
K. R. W. Jones,
Phys. Rev. A \textbf{50}, 3682 (1994).

\bibitem{Massar-Popescu1995}
S. Massar and S. Popescu,
Phys. Rev. Lett. \textbf{74}, 1259 (1995).

\bibitem{Derka-etal1998}
R. Derka, V. Bu\u{z}ek, and A. K. Ekert,
Phys. Rev. Lett. \textbf{80}, 1571 (1998).

\bibitem{Gill-Massar2000}
R. D. Gill and S. Massar,
Phys. Rev. A \textbf{61}, 042312 (2000).

\bibitem{Embacher-Narnhofer2004}
F. Embacher, and H. Narnhofer,
Annals of Physics \textbf{311}, 220 (2004).

\bibitem{deBurgh-etal2008}
M. D. de Burgh, N. K. Langford, A. C. Doherty, and A. Gilchrist,
Phys. Rev. A \textbf{78}, 052122 (2008).

\bibitem{Guta-etal2008}
M. Gu\c{t}\u{a}, B. Janssens, and J. Kahn,
Commun. Math. Phys. \textbf{277}, 127 (2008).

\bibitem{Fischer-etal2000}
D. G. Fischer, S. H. Kienle, and M. Freyberger,
Phys. Rev. A \textbf{61}, 032306 (2000).

\bibitem{Hannemann-etal2002}
Th. Hannemann, D. Reiss, Ch. Balzer, W. Neuhauser, P. E. Toschek, and Ch. Wunderlich,
Phys. Rev. A \textbf{65}, 050303(R) (2002).

\bibitem{Bagan-etal2005}
E. Bagan, A. Monras, and R. Mu\~{n}oz-Tapia,
Phys. Rev. A \textbf{71}, 062318 (2005).

\bibitem{Happ-Freybergerl2008}
C. J. Happ and M. Freyberger,
Phys. Rev. A \textbf{78}, 064303 (2008).

\bibitem{Kravtsov-etal2013}
K. S. Kravtsov, S. S. Straupe, I. V. Radchenko, N. M. T. Houlsby, F. Huszár, and S. P. Kulik,
Phys. Rev. A \textbf{87}, 062122 (2013).

\bibitem{Nielsen-Chuang2000}
M. A. Nielsen and I. L. Chuang, \textit{Quantum Computation and Quantum Information} (Cambridge University Press, Cambridge, UK, 2000).

\bibitem{Hwang2003}
W.-Y. Hwang,
Phys. Rev. Lett. \textbf{91}, 057901 (2003).

\bibitem{Lo-etal2005}
H.-K. Lo, X. Ma, and K. Chen,
Phys. Rev. Lett. \textbf{94}, 230504 (2005).

\bibitem{Shapiro2008}
J. H. Shapiro,
Phys. Rev. A \textbf{77}, 052330 (2008).

\bibitem{Barbosa-etal2003}
G. A. Barbosa, E. Corndorf, P. Kumar, and H. P. Yuen,
Phys. Rev. Lett. \textbf{90}, 227901 (2003).

\bibitem{Yuen2009}
H. P. Yuen,
IEEE J. Sel. Top. Quantum Electron. \textbf{15}, 1630 (2009).

\bibitem{Kak2006}
S. Kak,
Found. Phys. Lett. \textbf{19}, 293 (2006).

\bibitem{Chan-etal2015}
K. W. C. Chan, M. El Rifai, P. K. Verma, S. Kak, and Y. Chen,
International Journal on Cryptography and Information Security, Vol. 5, No. 3/4, pp. 1–13 (2015);
arXiv:1503.05793 [quant-ph].

\bibitem{Zhao-etal2007}
Y. Zhao, B. Qi and H.-K. Lo,
Appl. Phys. Lett. \textbf{90}, 044106 (2007).

\bibitem{Tang-etal2013}
Y.-L. Tang, H.-L. Yin, X. Ma, C.-H. F. Fung, Y. Liu, H.-L. Yong, T.-Y. Chen, C.-Z. Peng, Z.-B. Chen, and J.-W. Pan,
Phys. Rev. A \textbf{88}, 022308 (2013).

\bibitem{Bagan-etal2004}
E. Bagan, M. Baig, R. Mu\~{n}oz-Tapia, and A. Rodriguez,
Phys. Rev. A \textbf{69}, 010304(R) (2004).

\bibitem{Helstrom1976}
C. W. Helstrom,
\textit{Quantum Detection and Estimation Theory}
(Acadenic Press, New York, 1976),
Chapter 8.

\bibitem{Audenaert-etal2007}
K. M. R. Audenaert, J. Calsamiglia, R. Muñoz-Tapia, E. Bagan, Ll. Masanes, A. Acin, and F. Verstraete,
Phys. Rev. Lett. \textbf{98}, 160501 (2007).

\end{thebibliography}
\end{document}